\documentclass[a4paper]{article}

\usepackage{INTERSPEECH2022}
\usepackage{bm}
\usepackage{hyperref}
\usepackage{multirow}
\usepackage{multicol}
\usepackage{subcaption}
\usepackage{lipsum}
\usepackage{diagbox}
\usepackage{float}
\usepackage{array}
\usepackage{tabularx,ragged2e}
\usepackage{comment}
\newcommand{\etal}{\emph{et al.}\ }

\title{DDOS: A MOS Prediction Framework utilizing Domain Adaptive Pre-training and Distribution of Opinion Scores}
\name{Wei-Cheng Tseng, Wei-Tsung Kao, Hung-yi Lee}
\address{
  Graduate Institute of Communication Engineering, National Taiwan University}
\email{r09942094@ntu.edu.tw, r09942067@ntu.edu.tw, hungyilee@ntu.edu.tw}

\begin{document}

\maketitle

\begin{figure*}[!h]
    \centering
    \includegraphics[width=0.915\linewidth]{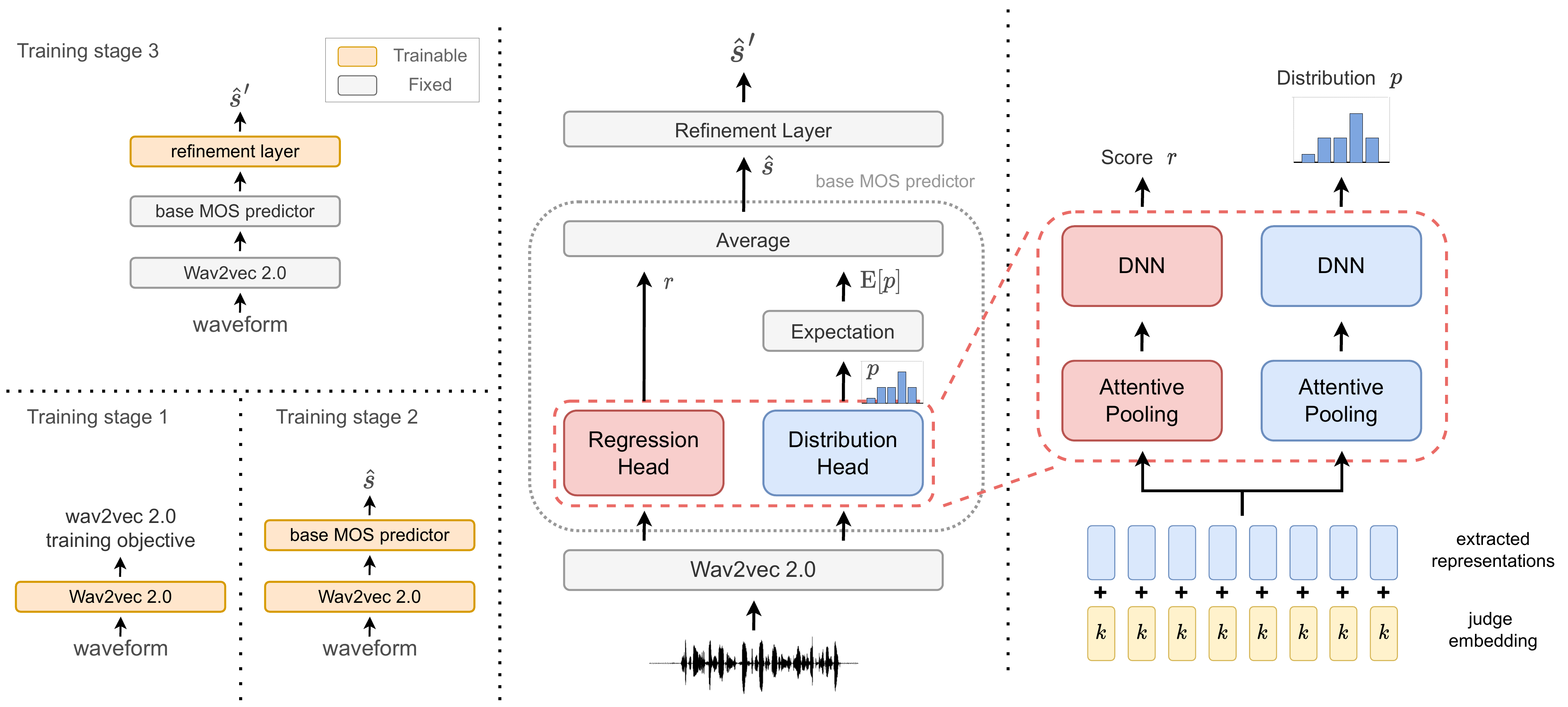}%
    \caption{The illustration of the proposed framework. DDOS is composed of wav2vec 2.0, the base MOS predictor module with two prediction heads, and the refinement layer. In the first training stage, we continue pre-training wav2vec 2.0 on MOS datasets. Second, we fine-tune wav2vec 2.0 and the base MOS predictor. Finally, the refinement layer is optimized in the third training stage.}
    \label{fig:framework}
\end{figure*}

\begin{abstract}
  Mean opinion score (MOS) is a typical subjective evaluation metric for speech synthesis systems. Since collecting MOS is time-consuming, it would be desirable if there are accurate MOS prediction models for automatic evaluation. In this work, we propose DDOS, a novel MOS prediction model. DDOS utilizes domain-adaptive pre-training to further pre-train self-supervised learning models on synthetic speech. And a proposed module is added to model the opinion score distribution of each utterance. With the proposed components, DDOS outperforms previous works on BVCC dataset. And the zero-shot transfer result on BC2019 dataset is significantly improved. DDOS also wins second place in Interspeech 2022 VoiceMOS challenge in terms of system-level score.
\end{abstract}

\noindent\textbf{Index Terms}: MOS predicion, self-supervised learning

\section{Introduction}
\label{sec:intro}
Mean Opinion Score (MOS) is a widely used evaluation metric in speech synthesis. For applications such as text-to-speech (TTS) and voice conversion (VC), there is usually a lack of ground truth. People thus resort to human evaluation to monitor the quality of their speech synthesis systems. An opinion score to a speech utterance is an integer given by human judges, in the range of 1 to 5. A Higher opinion score means the judge considers the speech has better quality and vice versa. MOS is the average of opinion scores over several judges. Collecting opinion scores is time-consuming and costly since it often involves many human judges and spends a lot of time listening to the synthesized speech. Assume there is a total of $N$ judges. A common scenario is that for each utterance, people randomly sample $K$ ($\ll N$) judges to collect their opinion scores and calculate MOS but not let every judge listen to the same utterances.

Due to the cost of collecting opinion scores, several deep-learning-based methods have been proposed to predict the MOS. Lo \etal\cite{lo2019mosnet} first modeled the MOS prediction as a regression problem and proposed MOSNet, a CNN-BiLSTM model, to predict MOS. Leng \etal\cite{leng2021mbnet} proposed MBNet, in which the authors added the regression modeling on opinion scores given by each judge to the model. They found that utilizing judge information improves generalizability. Huang \etal\cite{huang2021ldnet} proposed LDNet to improve the model structure of MBNet. LDNet merges the MOS prediction network and the judge-dependent part in MBNet into a single encoder-decoder network. Cooper \etal\cite{cooper2021generalization} improved MOSNet by two data augmentations, changing speed and adding silence. 

These works only use labeled datasets to train. However, collecting a large-scale MOS dataset is difficult. The size of datasets could limit the performance. Therefore, researchers start to incorporate unsupervised learning into MOS prediction. Self-supervised learning (SSL) is a powerful unsupervised learning technique. Pre-trained on large-scale speech-only corpora, the SSL models can extract general-purpose representations that benefit a wide variety of tasks including ASR, speaker verification, etc~\cite{yang2021superb}. For MOS prediction, Tseng \etal\cite{tseng21b_interspeech} evaluated four SSL models, APC~\cite{chung2019unsupervised}, CPC~\cite{oord2018representation}, TERA~\cite{liu2020tera}, and Wav2vec2~\cite{baevski2020wav2vec}. The authors concluded that Wav2vec2 is the best among these models. They also showed that before fine-tuning on labeled datasets, the representations from Wav2vec2 are discriminative between high-quality and low-quality speech. 
MOSA-Net~\cite{zezario2021deep} originally used spectral and temporal features along with HuBERT~\cite{hsu2021hubert} representations to predict the quality of noisy speech. 
Recently, it is shown to be effective in predicting MOS.
Cooper \etal\cite{cooper2021generalization} studied the generalizability of SSL models. The SSL models were fine-tuned on English TTS and VC MOS datasets and transferred to Chinese and Japanese speech. The experiment results showed that SSL models generalize well even under the zero-shot scenario. 

In this work, we would like to further improve SSL-based MOS prediction models. We notice that recent SSL models are pre-trained on natural speech corpus such as LibriSpeech~\cite{LbirSpeech}, while encountering synthetic speech in MOS prediction. This domain mismatch could affect the quality of the representations. Moreover, we find that the opinion score distribution of each utterance is not explicitly used in the training procedure of previous works. Based on these observations, our contributions include:
\begin{itemize}
    \item We propose applying domain-adaptive pre-training before fine-tuning on the labeled MOS datasets to reduce domain mismatch between speech in the pre-training corpus and fine-tuning corpus.
    
    \item We design a sub-module to make use of opinion score distribution information explicitly. 
    
    \item The proposed model outperforms previous SSL-based MOS prediction models and non-SSL-based state-of-the-art model, LDNet.
\end{itemize}

\section{Proposed Method}
\label{sec:method}

\subsection{Overview of model structure}

The MOS prediction model structure is shown in the middle of Figure~\ref{fig:framework}. 
We first use Wav2vec 2.0~\cite{baevski2020wav2vec} to encode the input waveform into a sequence of hidden representations. 
Judge ids are transformed to embeddings by a learnable embedding table and added to each of the hidden representations. 
The corresponding opinion score of 0 judge id is MOS. 
Then a base MOS predictor consisting of two sub-modules, regression head and distribution head, is added on top of Wav2vec 2.0.
The base MOS predictor takes the hidden representations as input and outputs a predicted MOS score $\hat{s}$.
The structure of the regression head and distribution head is shown on the right of Figure~\ref{fig:framework}. In each of these heads, an attentive pooling layer learns the importance of each representation and merges the sequence into a single representation. 
Then a 3-layer DNN transforms the representation into a predicted opinion score (regression head) or an opinion score distribution (distribution head). 
The expectation of the distribution represents predicted MOS. 
We average scores from the two heads as the final result of the base MOS predictor.
Finally, a refinement layer takes $\hat{s}$ as input and outputs an adjusted score $\hat{s}'$ by refining the scale of $\hat{s}$.

There are three training stages for our MOS prediction model as shown in the left of Figure~\ref{fig:framework}. 
In the first stage, we continue pre-training wav2vec 2.0 on the speech in MOS datasets, which is called domain-adaptive pre-training (DAPT). We detail DAPT in section~\ref{sec:DAPT};
in the second stage, we fine-tune wav2vec 2.0 and the base MOS predictor without the refinement layer; 
in the third stage, we fix the wav2vec 2.0 and the base MOS predictor and train the refinement layer.
We will detail the training procedure of the heads in section~\ref{sec:dis_head} and the one of the refinement layer in section~\ref{sec:refine}.

\subsection{Domain adaptive pre-training}
\label{sec:DAPT}
In MOS prediction, we need to predict the MOS of the synthetic utterances generated by other systems such as voice conversion or text-to-speech. 
While wav2vec 2.0 is pre-trained on real speech, i.e., LibriSpeech. 
This domain mismatch could limit the performance of the pre-trained models. 
So we load the wav2vec 2.0 pre-trained on LibriSpeech and continue pre-training it on the synthetic speech in the MOS dataset to deal with the domain mismatch. 
The procedure of pre-training on domain-specific datasets is called domain-adaptive pre-training (DAPT) in NLP literature~\cite{gururangan2020don}.
Recent works have also shown DAPT is effective in speech recognition~\cite{hsu2021robust}. 
To our best knowledge, this work is the first work trying to apply DAPT to MOS prediction.

\subsection{Regression head and distribution head}
\label{sec:dis_head}
Recently proposed MOS prediction models are usually trained on the datasets in which not only MOS $\Bar{s}$ is provided but also the opinion scores $\{s_i\}_{i=1}^K$ from each judge are available. 
Previous works conduct regression on these scores by minimizing $L_1$ or $L_2$ distance. The regression head in our base MOS predictor follows this practice. When the 0 judge id is input, the regression head minimizes the mean square error (MSE) between its output $r$ and $\Bar{s}$; when a specific judge id $i$ is input, it minimizes the MSE between $r$ and $s_i$.

Although it is intuitive to view the MOS prediction as a regression problem, the opinion score distribution of each utterance (over $\{s_i\}$) is not explicitly utilized in this training loss. 
We consider that the opinion score distribution of an utterance can also reflect its quality. 
Besides, as mentioned in section~\ref{sec:intro}, collecting MOS datasets is costly and time-consuming. We would like to use as much information in the datasets as possible.
Therefore, we propose distribution head to inject this information into our MOS prediction model.
The distribution head outputs a distribution $p$ over $\{1, 2, 3, 4, 5\}$. We use cross-entropy to train this head. When a non-zero judge id is provided, we let the target of $p$ be a one-hot vector with the $s_i$-th dimension equal to 1 and the other dimensions equal to 0 since $\forall i, s_i \in \{1, 2, 3, 4, 5\}$. When the 0 judge id is input, the target of $p$ is the opinion score distribution of the utterance
\begin{align}
    \frac{1}{K}(\sum_{t=1}^{K}\mathrm{1}_{\{s_t=1\}}, \sum_{t=1}^{K}\mathrm{1}_{\{s_t=2\}}, ..., \sum_{t=1}^{K}\mathrm{1}_{\{s_t=5\}}),
\end{align}
where $\mathrm{1}_{\{\cdot\}}$ is the indicator function. 
During inference, the expectation $\text{E}[p]$ is viewed as the predicted MOS of the distribution head. 
The average of $\text{E}[p]$ and $r$ is the final output of the base MOS predictor. 
We always input the 0 judge id to obtain predicted MOSs.

\begin{table*}[!t]
    \centering
    \caption{The performances of our framework and baselines on BVCC test set.}
        \begin{tabular}{l||r|r|r|r|r|r|r|r}
            \hline
             \multirow{2}{*}{} & \multicolumn{4}{c|}{Utterance-level} & \multicolumn{4}{c}{System-level} \\
             \cline{2-9}
             & MSE $\downarrow$ & LCC $\uparrow$ & SRCC $\uparrow$ & KTAU $\uparrow$ & MSE $\downarrow$ & LCC $\uparrow$ & SRCC $\uparrow$ & KTAU $\uparrow$ \\
             \hline \hline
             LDNet \cite{huang2021ldnet} & 0.334 & 0.787 & 0.785 & 0.599 & 0.178 & 0.873 & 0.873 & 0.691 \\
             MOSA-Net \cite{zezario2021deep} & 0.309 & 0.809 & 0.811 & 0.621 & 0.162 & 0.899 & 0.900 & 0.729\\
             SSL-MOS \cite{cooper2021generalization} & 0.277 & 0.869 & 0.869 & 0.690 & 0.148 & 0.924 & 0.921 & 0.769\\
             \hline
             DDOS (ours) & 0.212 & \textbf{0.880} & \textbf{0.880} & \textbf{0.707} & 0.110 & \textbf{0.933} & \textbf{0.932} & \textbf{0.782} \\
             
             \quad -\ DNN in head & 0.208 & 0.878 & 0.879 & 0.706 & 0.116 & 0.927 & 0.927 & 0.774 \\
             \quad -\ Distribution Head & 0.205 & 0.878 & 0.876 & 0.702 & 0.105 & 0.930 & 0.928 & 0.774 \\
             \quad -\ Regression Head & 0.221 & 0.877 & 0.878 & 0.704 & 0.115 & 0.932 & 0.931 & 0.779 \\
             \quad -\ refinement layer & 0.230 & 0.880 & 0.880 & 0.707 & 0.106 & 0.933 & 0.932 & 0.782 \\
             \quad -\ DAPT & \textbf{0.201} & 0.877 & 0.875 & 0.701 & \textbf{0.102} & 0.928 & 0.926 & 0.770 \\
             \quad -\ data augmentation &0.213 & 0.880 & 0.879 & 0.705 & 0.118 & 0.927 & 0.925 & 0.774\\
             \hline
        \end{tabular}
    
    \label{tab:bvcc}
\end{table*}

\subsection{Refinement Layer}
\label{sec:refine}
The base MOS predictor can achieve satisfying LCC and SRCC, but the MSE is still not acceptable. Accordingly, we add a refinement layer and try to reduce MSE. After the second stage of training, we infer the base MOS predictor on the whole training dataset and get the predicted MOSs. Then a linear regression model is trained to minimize the $L_2$ distance between the predicted MOSs and the ground truths. We choose a linear model to ensure that correlation metrics remain unchanged (as long as the weight of the linear model is positive).

At first glance, adding a refinement layer seems equivalent to adding one linear layer into the MOS prediction model. But they are subtly different: (1) We do not jointly train the refinement layer with the MOS prediction model. (2) The refinement layer is optimized by the whole training dataset at once and has a closed-form solution. We do not apply joint training because we think that our large MSE stems from batch training, which only optimizes the MSE of part of the training data during each update. If we directly add a linear layer and jointly train, the linear layer would still access a batch of training data at once. Instead, the two-stage learning allows optimization of the whole training dataset. With this simple refinement, we can adjust the scale of our model's outputs, reduce MSE effectively, and do not degrade the performance on the other metrics.

\subsection{Data augmentation}
Cooper \etal\cite{cooper2021generalization} show that adding silence and changing speed as data augmentations can improve MOSNet while seems not helpful for SSL-based MOS prediction models.
Although the authors do not explain the motivation for choosing these augmentations, we consider that these augmentations do not influence MOS.
Inspired by the previous work, we adopt several data augmentations to investigate whether they are beneficial to DDOS or not, including speed-up or slow-down the audio, tuning the pitch of the human voice, and changing the tempo of the audio (without affecting the pitch).
We exclude adding or removing silence since we find it is not helpful in a preliminary study.
Typical data augmentation such as SpecAugment~\cite{park19e_interspeech} is not included since this kind of data augmentation affects speech quality and is unreasonable for MOS prediction.



\section{Experiments Setup}
\label{sec:exp}
In this paper, two datasets are involved in the experiments: \textbf{BVCC}~\cite{cooper2021generalization} and \textbf{BC2019}~\cite{wu2019blizzard}. 
BVCC is a newly collected MOS dataset that contains 7106 English samples from previous Blizzard Challenge for TTS~\cite{king2008blizzard,kinga2009blizzard,King2010TheBC, King2011TheBC,King2014TheBC} and Voice Conversion Challenge~\cite{Toda2016TheVC,d8e0d7191cac4700bc1dacd63bd43196,lorenzo2018voice,zhao2020voice,Das2020} as well as synthesized samples from systems implemented in ESPNet~\cite{watanabe2018espnet}.
Samples in this dataset are re-evaluated with one unified listening test and divided into training/development/test splits with 70\%/15\%/15\% ratios. 
BC2019 contains Chinese TTS samples submitted to 2019 Blizzard Challenge~\cite{wu2019blizzard}, where each sample is rated by 10 to 17 judges. 
We follow the setup in Interspeech 2022 VoiceMOS challenge~\cite{huang2022voicemos} to create training/development/test splits with 136/136/540 samples, respectively, along with an unlabeled set that contains 540 samples without rating.
Four evaluation metrics are used in the experiments: mean squared error (MSE), linear correlation coefficient (LCC), Spearman’s rank correlation coefficient (SRCC), and Kendall rank correlation coefficient (KTAU). MSE measures the absolute difference between predicted scores and ground truths, while the rest tell how correlated they are.

Three baseline models are included in the experiments for comparison: LDNet~\cite{huang2021ldnet}, MOSA-Net~\cite{zezario2021deep}, SSL-MOS~\cite{cooper2021generalization}. We use the checkpoints provided by their official implementation repositories.
For DDOS, we continue to pre-train wav2vec 2.0-base model on the training split of both datasets along with the unlabeled set of BC2019 for 100 epochs with a batch size of 256.
Then we apply data augmentations for BVCC training set using the \emph{sox} toolkit\footnote{http://sox.sourceforge.net}, resulting in a total of 7 times the original data.
The pre-trained model is fine-tuned on these data for 20k steps with a learning rate of 1e-5. 
We apply warm-up in the first 500 steps and linearly decay the learning rate in the remaining steps.
Validation is performed every 250 steps, and the checkpoint with minimum validation loss is then used for evaluation. 

Apart from evaluating on BVCC, we investigate the transferability of DDOS (fine-tuned on BVCC) by testing it on BC2019.
There are three evaluation setups: (1) zero-shot: the zero-shot setting does not allow using any labels in BC2019 for training. 
We directly test DDOS and baselines on BC2019 after being fine-tuned on BVCC. We remove the refinement layer in this setup as the data distributions of BVCC and BC2019 are different. 
(2) few-shot: under few-shot setting, 10 labeled utterances from BC2019 are available for training. 
We fine-tuned DDOS and baselines on these labeled data for 50 epochs before testing. 
(3) full dataset: we use all 136 utterances in BC2019 to fine-tune DDOS and baselines. We train DDOS for 50 epochs and baselines following their official configurations. 
All the experiment results are averages of three independent runs.

\begin{table*}[!t]
    \centering
    \caption{The utterance-level performance (upper half) and the system-level score (lower half) of our framework along with four baselines on BC2019 test set. We compare all the systems in three different setups: zero-shot, few-shot and full dataset.}
    \begin{tabularx}{\textwidth}{
    @{}
    l||c|c|c|c|c|c|c|c|c|c|c|c
    @{}
    }
        \hline
         \multirow{2}{*}{\footnotesize Utterance-level} & \multicolumn{4}{c|}{zero-shot} & \multicolumn{4}{c|}{few-shot} & \multicolumn{4}{c}{full dataset} \\
         \cline{2-13}
          & \footnotesize MSE$\downarrow$& \footnotesize LCC$\uparrow$& \footnotesize SRCC$\uparrow$& \footnotesize KTAU$\uparrow$& \footnotesize MSE$\downarrow$& \footnotesize LCC$\uparrow$& \footnotesize SRCC$\uparrow$& \footnotesize KTAU$\uparrow$& \footnotesize MSE$\downarrow$& \footnotesize LCC$\uparrow$& \footnotesize SRCC$\uparrow$& \footnotesize KTAU$\uparrow$ \\
         \hline \hline
         \footnotesize LDNet \cite{huang2021ldnet} & 1.587 & 0.384 & 0.365 & 0.252 & 0.994 & 0.530 & 0.509 & 0.354 & 0.287 & 0.844 & 0.787 & 0.592  \\
         \footnotesize MOSA-Net \cite{zezario2021deep} & 3.460 & 0.316 & 0.401 & 0.281 & 0.690 & 0.553 & 0.566 & 0.406 & 0.284 & 0.854 & 0.806 & 0.616 \\
         \footnotesize SSL-MOS \cite{cooper2021generalization} & 3.924 & 0.394 & 0.464 & 0.345 & 0.476 & 0.728 & 0.692 & 0.508 & 0.260 & 0.888 & 0.849 & 0.664  \\
         \hline
         \footnotesize DDOS (ours) & 1.331 & 0.678 & 0.694 & 0.502 & 0.338 & 0.812 & 0.751 & 0.569 & 0.169 & 0.914 & 0.887 & 0.710  \\
         \hline \hline
         \multirow{2}{*}{System-level} & \multicolumn{4}{c|}{zero-shot} & \multicolumn{4}{c|}{few-shot} & \multicolumn{4}{c}{full dataset} \\
         \cline{2-13}
          & \footnotesize MSE$\downarrow$& \footnotesize LCC$\uparrow$& \footnotesize SRCC$\uparrow$& \footnotesize KTAU$\uparrow$& \footnotesize MSE$\downarrow$& \footnotesize LCC$\uparrow$& \footnotesize SRCC$\uparrow$& \footnotesize KTAU$\uparrow$& \footnotesize MSE$\downarrow$& \footnotesize LCC$\uparrow$& \footnotesize SRCC$\uparrow$& \footnotesize KTAU$\uparrow$ \\
         \hline \hline
         \footnotesize LDNet \cite{huang2021ldnet} & 1.307 & 0.500 & 0.473 & 0.354 & 0.789 & 0.674 & 0.710 & 0.539 & 0.091 & 0.952 & 0.934 & 0.791 \\
         \footnotesize MOSA-Net \cite{zezario2021deep} & 3.259 & 0.203 & 0.495 & 0.422 & 0.633 & 0.526 & 0.564 & 0.434 & 0.071 & 0.967 & 0.954 & 0.846 \\
         \footnotesize SSL-MOS \cite{cooper2021generalization} & 3.674 & 0.402 & 0.514 & 0.422 & 0.314 & 0.854 & 0.820 & 0.647 & 0.093 & 0.971 & 0.975 & 0.889  \\
         \hline
         \footnotesize DDOS (ours) & 1.119 & 0.766 & 0.797 & 0.637 & 0.185 & 0.895 & 0.838 & 0.705 & 0.052 & 0.976 & 0.955 & 0.848 \\
         \hline
    \end{tabularx}
    \label{tab:bc_sys}
\end{table*}

\begin{table}[!t]
    \centering
    \caption{Zero-shot ablation study.}
        \begin{tabular}{l||c|c|c|c}
            \hline
             System-level & \footnotesize MSE$\downarrow$ &\footnotesize LCC$\uparrow$ &\footnotesize SRCC$\uparrow$ &\footnotesize KTAU$\uparrow$ \\
             \hline \hline
             \footnotesize DDOS & 1.119 & 0.766 & 0.797 & 0.637\\
             \footnotesize -\ DNN in head & 0.896 & 0.793 & 0.822 & 0.668 \\
             \footnotesize -\ Distribution Head & 0.921 & 0.755 & 0.779 & 0.614 \\
             \footnotesize -\ Regression Head & 4.682 & 0.773 & 0.804 & 0.645 \\
             \footnotesize -\ DAPT & 2.813 & 0.696 & 0.704 & 0.559 \\
             \footnotesize -\ data augmentation & 0.948 & 0.813 & 0.813 & 0.660 \\
             \hline
        \end{tabular}
    \label{tab:zero_ablation}
\end{table}

\begin{table}[t]
    \centering
    \caption{The ranking of DDOS among 22 participants in Interspeech 2022 VoiceMOS challenge.}
        \begin{tabular}{l||r|r|r|r}
            \hline
             Ranking & MSE & LCC & SRCC & KTAU \\
             \hline \hline
             utterance & 4 & 4 & 5 & 5 \\
             \hline
             system & 2 & 2 & 2 & 3 \\
             \hline
        \end{tabular}
    \label{tab:voicemos}
\end{table}

\section{Results}
\label{sec:exp_result}

\subsection{Quantitative Results}
\label{sec:bvcc_result}
We first evaluate the performance of our framework along with baselines on BVCC test set.
Table \ref{tab:bvcc} lists the performance of the proposed framework along with three baselines.
Our framework surpasses all previous works in terms of all correlation coefficient metrics with a considerable amount while greatly reducing MSE at both utterance-level and system-level. 
We then inspect the contribution of each module to the performance. The results are shown in the lower half of Table~\ref{tab:bvcc}. We first analyze the model structure and then the training algorithms. First, we replace the DNN in the heads with a linear layer (- DNN head) to check whether the improvement comes from more parameters. The result of "- DNN head" is still better than SSL-MOS, which eliminates this concern. Second, we study the effect of modeling score distribution. 
The row "- Distribution head" shows that when we remove this modeling component, both system-level score and utterance-level score degrade. However, we can not use the distribution head alone. Without regression head, the utterance-level scores drop, and MSE inclines. 
Next, we find that the refinement layer effectively reduces utterance-level MSE for DDOS. 
Finally, we would like to confirm whether DAPT and data augmentation are helpful or not. When we exclude DAPT, all correlation metrics decline significantly, which implies that DAPT is necessary for DDOS. For data augmentation, although it makes a negligible difference in utterance-level scores, it enhances system-level scores. Thus, applying these kinds of data augmentation is also helpful.

\subsection{Transferability}
\label{sec:generalize}

In this section, we examine the transferability of DDOS by fine-tuning it on BVCC training set and evaluating it on BC2019. 
The utterance-level and system-level results are shown in the upper half and the lower half of Table~\ref{tab:bc_sys}, respectively. 
For the zero-shot setting, SSL-MOS and MOSA-Net are worse than LDNet in terms of MSE and system-level LCC but better than LDNet in terms of SRCC and KTAU.
So simply incorporating SSL models does not guarantee transferability in all metrics. 
DDOS exceeds all baselines considerably, which shows that DDOS is more robust and generalizes better than previous works. 
Table~\ref{tab:zero_ablation} shows the contribution of DDOS's components to the transferability by an ablation study. 
We only show system-level results since the trends of utterance-level results are similar.
DAPT is critical for zero-shot transfer. 
Distribution head also leads to improvement in correlation metrics.  
Instead, adding DNN in heads and data augmentation is detrimental. 

As for the few-shot setting, DDOS also outperforms the baselines significantly. 
Meanwhile, compared to the zero-shot setting, the performances of all systems are also boosted obviously. 
These results suggest that it is economical for us to conduct a relatively small-scale MOS test and use the collected data to fine-tune MOS prediction models before applying them to unseen datasets. 
Last, when the full dataset is available, all the models can achieve satisfactory system-level performance. 
DDOS is competitive with SSL-MOS in terms of system-level score. 
But when it comes to utterance-level scores, DDOS again provides remarkable improvements. 
Consequently, we can conclude that the proposed methods mainly intensify the generalization ability of the MOS prediction model.

\subsection{Interspeech 2022 VoiceMOS Challenge results}

We submit DDOS to Interspeech 2022 VoiceMOS Challenge. The final ranking is listed in Table~\ref{tab:voicemos}. Among 22 participants, DDOS gets fourth/fifth place in the utterance-level evaluation and wins second place in system-level evaluation in the main track (BVCC testing set). So DDOS is suitable for evaluating system-level MOS. 
We would try to modify DAPT to obtain better utterance-level results for future works since we observe that removing DAPT slightly improves utterance-level MSE.

\section{Conclusions}

This paper proposes a new MOS prediction model, DDOS, that utilizes domain-adaptive pre-training and models the opinion score distribution of each utterance to improve generalizability. DDOS obtains better performance than previous methods based on self-supervised learning models. Furthermore, it substantially improves the zero-shot and few-shot transferability. DDOS also wins second place in Interspeech 2022 VoiceMOS challenge main track in terms of system-level score, which verifies that DDOS is suitable for evaluating system-level MOS.

\section{Acknowledgement}

We thank to National Center for High-performance Computing (NCHC) of National Applied Research Laboratories (NARLabs) in Taiwan for providing computational and storage resources.

\bibliographystyle{IEEEtran}

\bibliography{mybib}


\end{document}